\documentstyle[aps,prd,amssymb,eqsecnum,epsf,twocolumn]{revtex}
\newcommand{\ba}{\begin{array}}
\newcommand{\ea}{\end{array}}
\newcommand{\be}{\begin{equation}}
\newcommand{\ee}{\end{equation}}
\newcommand{\bea}{\begin{eqnarray}}
\newcommand{\eea}{\end{eqnarray}}
\newcommand{\nn}{\nonumber}

\newcommand{\newc}{\newcommand}
\newc{\ra}{\rightarrow}
\newc{\lra}{\leftrightarrow}
\newc{\beq}{\begin{equation}}
\newc{\eeq}{\end{equation}}
\newskip\humongous \humongous=0pt plus 1000pt minus 1000pt

\newif\ifdtup

\relax
\title{Rotating Toroidal  Branes in Supermembrane and Matrix Theory}
\author{M. Axenides$^a$, E. G. Floratos$^a$,$^b$, L.
Perivolaropoulos$^a$,$^c$ }
\address{$^a$ Institute of Nuclear Physics, N.C.S.R. Demokritos,  153
10, Athens, Greece, \\e-mail:minos@inp.demokritos.gr,\\ $^b$
National University of Athens, Athens, Greece,
e-mail:manolis@inp.demokritos.gr \\$^c$ Physics Department,
University of Patras, Patras, Greece,
e-maill:leandros@physics.upatras.gr\thanks{Present Address}}

\date{\today}
\draft

\begin{document}

\maketitle

\begin{abstract}

In the lightcone frame, where the supermembrane theory and the
Matrix model are strikingly similar, the equations of motion admit
an elegant complexification in even dimensional spaces. Although
the explicit rotational symmetry of the target space is lost, the
remaining unitary symmetries apart from providing a simple and
unifying description of all known solutions suggest new ones for
rotating spherical and toroidal membranes. In this framework the
angular momentum is represented by $U(1)$ charges which balance
the nonlinear attractive forces of the membrane. We examine in
detail a six dimensional rotating toroidal membrane solution which
lives in a $3$-torus, $T^3$ and admits stable radial modes. In
Matrix Theory it corresponds to a toroidal N-$D_{0}$ brane bound
state.  We demonstrate its existence and discuss its radial
stability.
\end{abstract}


\section{Introduction}
String Theory and the more speculative nonperurbative version of
it, M-theory, still is the only surviving candidate for the
unification of gravity with quantum mechanics and the other
fundamental gauge interactions. The D-branes of string theory are
the long searched for gravitational instantons and can be
described together with their dualities by simple geometrical
symmetries of compactified $M_{2}$ and $M_{5}$ branes. The problem
for their interpretation as true gravitational solitons is that
they are usually static and possess infinite mass and charge if
not wraped in compact submanifolds.

The fundamental degrees seem to be the $D_{0}$ branes and many
attempts have been made to construct all the others as $D_{0}$
bound states. On the other hand in the Matrix model, bound states
of $D_{0}$ branes with spherical topologies having finite mass,
charge and angular momentum have been constructed either via
background field improved Matrix model or by introducing
rotational degrees of freedom. No supergravity solutions dual to
these bound states have been constructed, although their existence
is considered to be a valid hypothesis.

We adopt, as a conjecture to be proved, that the quantum
supermembrane theory is identical with the M-theory. The true
nonsingular gravitational solitons of $11$-dimensional
supergravity we expect them to be the duals of finite mass, flux,
angular momentum solutions of supermembrane theory in flat eleven
dimensional spacetime. The above picture is true for the infinite
mass and charge $M_2$ and $M_5$ branes after compactification.
Another possible interpretation for the compact solutions of the
supermembrane theory in the lightcone frame and flat background is
that of PP waves,  i.e. infinite momentum boosts of analogous
supergravity solutions with the same symmetries.

From this point of view  the search for stable classical
supermembrane solutions which can be used as true saddle points of
the supermembrane path integral quantization, while particles and
strings are singular saddle points appears to be worthwhile.

In this work we continue our search for supermembrane solutions
introducing the complexification of the light cone frame equations
of motion in flat space. We show that this framework encompasses
all previously found spherical and toroidal rotating membranes and
more importantly it suggests new ones. As an example we present a
rotating toroidal membrane in six or eight dimensions ($C^3$ and
$C^4$) which lives in three or four dimensional tori and  the
angular momentum is represented by three or correspondingly four
charges which balance the nonlinear attractive forces of the
membrane. We also demonstrate its radial stability.

In chapter II we describe the complexification of the lightcone
membrane equations of motion in even dimensions and we present our
generalized ansatz for the factorization of time. As a consequence
we show that all previously known solutions in the literature can
be deduced as particular examples.

In chapter III we exhibit a new interesting class of solutions
which describe toroidal rotating membranes embedded in a $T^{3}$
complex torus. These solutions may be relevant for new
supersymmetric toroidal compactifications of the M-theory. We
demonstrate their stability against radial pertutbations.

In chapter IV we utilize the isomorphism between the
nonperturbative sectors of Matrix and Supermembrane Theories and
demonstrate the existence and radial stability of the
corresponding Toroidal N-$D_{0}$ brane solutions.

We close by presenting some concluding remarks and speculations
about the supergravity duals of our new configurations.

\section{Complex Dimensions and Rotating Membranes}
%Nicolai-Hoppe(Covariant formulation and Spherical and Toroidal
%Target Spaces), Hoppe( Matrix Eqs. for Spherical and Toroidal Matrix
%Branes), Taylor( Spherical Matrix Branes and Supergravity
%Potentials, velocity corrections and rotating matrix branes),
%Myers Stabilization, Rey(Gravitation Matrix Q-Balls),
%Savvidy-Hammarck and Sav-Sav...
%The equation of motion for the spherical supermembrane in six dimensions may
%be written as
The nonlinear attractive force as well as the absence of any
coupling constant for the membrane is known to be the main
difficulty  for the analysis of its dynamics at the classical and
quantum levels\cite{CT,Y}. In order to understand the excitation
spectrum of the membrane it was obvious that one has to find
stable rotating solutions or to change the topology of space and
wrap static membranes along topologically similar cycles.

In an interesting work by Nicolai and Hoppe\cite{NH} it became
clear that rotating solutions for closed spherical and toroidal
membranes exist in higher dimensional spherical spaces. With the
discovery of the supermembrane Lagrangian, the search for
supersymmetric membrane configuration in curved spacetime received
a lot of attention \cite{D}. Later on M-theory was invented as a
strong coupling limit of type II-A string theory\cite{TWSD} and
the Matrix model, a reincarnation of the $SU(N)$ trancation of the
supermembrane system, was proposed as a candidate for the M-theory
\cite{BFSS}. The Matrix model description of the string theory and
its strong coupling limit was analyzed in detail and mechanisms
for understanding the fundamental excitations were proposed. Today
we understand the matrix model classical solutions as $D_0$ bound
states which form various topologies, spherical toroidal and other
fuzzy configurations\cite{Tay,KT}. The last few years rotating
Matrix spherical and toroidal configurations were found
\cite{TRRH}. For the spherical topology more detailed
investigations uncovered a rich structure of moduli and stability
in certain cases was shown\cite{HSS,AFP}.

In this section we review the existing solutions in a unified
framework in the light cone frame and flat Euclidean spaces. The
Hamiltonian  for the bosonic part of the supermembrane using
Darboux parametrization of the membrane surface  (Area element
$dA=\;d\sigma_{1}d\sigma_{2}$) is:

\be\label{hamil} H\;=\;\frac{1}{8\pi T}\int d^{2}\sigma\left[
\frac{1}{2} \dot{X}_{i}^2
+\frac{1}{4}\left\{X_{i},X_{j}\right\}^{2}\right]\;\;i=1,\ldots,d\ee
where $T$ is the membrane tension. The corresponding equations of
motion are given by:

\begin{equation}
  \label{eq:motion}
  \ddot{X}_i= \left\{ X_j,\left\{X_j, X_i \right\}\right\} \; \; i,j=1,..,d
\end{equation}
where summation is implied in the $j$ indices and $\left\{
\right\}$ stands for the Poisson bracket with respect to the
angular coordinates $\sigma_{1},\sigma_{2}$. The Gauss constraint
that also needs to be satisfied is

\begin{equation}
  \label{eq:gauss}
 \left\{\dot{X}_i, X_i \right\}=0
\end{equation}

This constraint is preserved by the equations of motion and
therefore if it is initially obeyed, as is the case with what
follows, it will be obeyed at all times. If $d=2k$ we define
$Y_{i}\equiv X_{i+k}$ with $i=1,..,k$. The equations of motion are
\bea\label{eq:motion1}
  \ddot{X}_i &=& \left\{ X_j,\left\{X_j, X_i \right\}\right\}+
   \left\{ Y_j,\left\{Y_j, X_i \right\}\right\}\nn \\
  \ddot{Y}_i &=& \left\{ X_j,\left\{X_j, Y_i \right\}\right\}+
   \left\{ Y_j,\left\{Y_j, Y_i \right\}\right\}
\eea

It is now convenient to introduce a convenient notation of complex
coordinates $Z_1$, $Z_2$,.. $Z_k$ defined as

\be \label{cmplx-not} Z_i = X_i + i Y_i, \; \;\;\; i=1,..,k   \ee
These coordinates transform the target space $R^{2k}$ of membrane
solutions
 into $C^k$ simplifying in this way the field equations considerably.
 In the context of these complex
coordinates of (\ref{cmplx-not}) the Hamiltonian takes the
following form:

\bea\label{hamill}  H\; &=&\;\frac {1} {8 \pi T}\int d^{2} \sigma
\;\; \left[  \frac{1}{2} |\dot{\vec{Z}}|^{2}\; + \; \frac{1}{2}
|\partial_{1} \vec{Z} \times \partial_{2} \vec{Z}|^{2}+ \right.\\
 & & \left.\mbox{} \frac{1}{2} |\partial_{1} \vec{Z} \times
\partial_{2}
 \vec{Z}^{*}|^{2}\; \right] \nn \eea where
$(\partial_{1}\vec{Z}\times\partial_{2}\vec{Z})_{i}\;=\;\epsilon_{ijk}
\partial_{1}Z_{j}\partial_{2}Z_{k}\;=\;\frac{1}{2}\epsilon_{ijk}
\left\{Z_{j},Z_{k}\right\}$ and $(\partial_{\alpha}\vec{Z})_{i} =
\frac{\partial}{\partial \sigma_{\alpha}} Z_{i},\;\;\;\alpha=1,2$
and $i=1,2,3$. Similarly the equations of motion transform to :
\be
  \label{eq:mot-cmplx}
  \ddot{Z}_i =\ddot{X}_i + i  \ddot{Y}_i ={1\over 2} \left[ \left\{ Z_j^*,
  \left\{Z_j, Z_i \right\}\right\}+ \left\{ Z_j,\left\{Z_j^*, Z_i
  \right\}\right\}\right] \ee
while the constraint becomes

\be \label{constr-cmplx} \left\{\dot{Z_i^*}, Z_i
\right\}+\left\{\dot{Z_i}, Z_i^* \right\} =0\;\;\;i,j=1,..,k
 \ee

Although the target space rotational symmetry $O(2k)$ becomes less
obvious in the complexified equations, their explicit $U(k)$
symmetry will simplify the description of the known solutions and
it will facilitate the search for new ones. Below we exhibit the
most general solution with factorization of time. To begin with
our most general ansatz is :

 \be\label{nans} Z_{i}=\zeta_{ij}(t)
f_{j}(\sigma_{1},\sigma_{2})\;\;\;\; i,j= 1,2,.....,k\ee where
$\zeta_{ij}$ is a $k\times k$ non-singular complex matrix and
$f_{j}$ are k-complex linearly independent functions of membrane
parameters. This ansatz provides a factorization of time in the
equations of motion and the reparametrization constraint if the
following conditions for the matrix $\zeta$ and the functions$
f_{i}$ are satisfied,

\bea\label{neweq1} \zeta\dag \;\zeta \;&=&\; \eta _{D}\nn\\
\zeta\dag\;\dot{\zeta} - \dot{\zeta}\dag\;\zeta\;&=&\; i\theta_{D}
\eea where  $\eta_{D}$ is a positive real diagonal matrix and
$\theta_{D}$ is a real diagonal matrix. Here we assume also that
$\left\{f_{i}^{\star}, f_{i}\right\}=0$ for every $i=1,...,k$. It
is not difficult to show that the above conditions for the matrix
$\zeta$, which are sufficient but not necessary, imply that it
must be complex diagonal up to multiplication on the left by  a
constant unitary matrix. We therefore adopt as an ansatz the
diagonal form:

\be\label{eq17} Z_i\;=\;\zeta_{i}(t)f_{i}\ee

It automatically satisfies the constraint and the NASC for the
factorization of time is given by

\be\label{eq18} \frac{1}{2}\left\{
f_j,\left\{f_j^*,f_i\right\}\right\}+\frac{1}{2}\left\{ f_j^*,
\left\{f_j,f_i\right\}\right\}= - \nu_{ji}\; f_{i}\ee for every
$i$ and $j$. The time dependent complex scale factors $\zeta_{i}$
satisfy

\be\label{eq19} \ddot{\zeta_i}\;\;=\;\;
-(\sum_{j}\;\;|\zeta_j|^2\nu_{ji})\zeta_{i}\ee

Firstly we treat the case of spherical membranes and we choose
$f_i\;=\;e_i,\;\;i=1,2,3 $, where

\be\label{eq5}e_{1}=\sin\theta \; \cos\phi,\;\; e_{2}=\sin\theta
\; \sin\phi,\;\; e_{3}=\cos\theta\ee

 As a consequence
 $\nu_{ij}=\delta_{jj}-\delta_{ij}$
 and the equations of motion (\ref{eq19}) become

 \be\label{eq20}\ddot{\zeta_i}\;=\; - \sum_{j} |\zeta_j|^2
(\delta_{jj}-\delta_{ij} )\zeta_i\;\;=\;\;-(|\vec{\zeta}|^2 -
|\zeta_i|^2 ) \zeta_i \ee

In order to make explicit the conserved
 quantities, we separate the moduli and the phases

\be\label{eq21}\zeta_i\;\;=\;\; \lambda_{i}(t) e^{i
\chi_{i}(t)}\ee

It is obvious that there are three $U(1)$ conserved charges

\be\label{eq22} Q_i\;\;=\;\; -\frac{i}{2}( \dot{\zeta_i}
\zeta_{i}^{*} - \dot{\zeta_i}^{*} \zeta_{i})\ee which in turn
implies

\be\label{eq23} \dot{\chi_i}\;\;=\;\; \frac{Q_i}{\lambda_i^2}\ee

By using (\ref{eq23}) we find

\be\label{eq24}\ddot{\lambda}_i -\frac{Q_i^2}{\lambda_i^3} \;+\;
(\vec{\lambda}^2 - \lambda_i^2) \lambda_i =0  \ee and get the
conserved energy of the system to be

 \be\label{energy}E= \sum_{i} \frac{1}{2} \dot{\lambda_{i}}^{2}
 +\frac{1}{2}\sum_{i} \frac{Q_{i}^{2}}{\lambda_{i}^{2}} +
 \frac{1}{2}
 (\lambda_{1}^{2}\lambda_{2}^{2}+\lambda_{2}^{2}\lambda_{3}^{2}+
 \lambda_{3}^{2}\lambda_{1}^{2}) \ee

For a similar system see \cite{HSS}. We now proceed to show that
the Nicolai-Hoppe spherical membrane solution\cite{NH} arises as a
special case of the framework we have introduced. Indeed the
Nicolai-Hoppe ansatz in the lightcone frame can be written as

\be\label{eq1} Z_{i}\;=\; \lambda(t)\;\; (Uf)_{i}\ee where the
unitary $3 \times 3$ matrix U is an overall time dependent phase
 \be\label{eq2} U\;=\; e^{i\chi(t)I} \;\;\;\;\;\;\in
U(3)\ee with I being identified as the identity $3\times3$ matrix.
In this case all scale factors $\lambda_i$ as well as all phases
are equal.
% \be\label{eq3}  A^\dagger\;=\; A
%\ee
%For the rotating spherical membrane the ansatz becomes
%\be\label{eq4}
%Z_{i}{t,\theta,\phi}\;=\; \zeta_{i}(t) e_{i}
%\ee
%and $e_{i}$ define the embedding in the space $C^{3}$.
The factorization of time for the three functions $f_{i}$ imposes
the condition

\be\label{eq3}\left\{
f_{k}^{\star},\left\{f_{k},f_{i}\right\}\right\} + \left\{f_{k},
\left\{f_{k}^{\star},f_{i}\right\}\right\}= -2 \nu f_{i}\ee where
$\nu$ is obviously positive and summation over the index $k$ is
implied. Nicolai and Hoppe choose $f_i\;=\;e_i$ (\ref{eq5}) which
gives $\nu=2$. With these assumptions we find,

\be\label{eq6} \ddot{\lambda}\;-\lambda\dot{\chi}^{2} \;\;
+\;\nu\lambda^{3}\;=\;0\ee

 \be\label{eq7}
\lambda\ddot{x}\;+\;2\dot{\lambda}\dot{x}\;=\;0\ee

They imply

\be\label{eq8} \dot{x}\;=\; \frac{L}{\lambda^2}\ee

\be\label{eq9}\; \ddot{\lambda}\;-\; \frac{L^2}{\lambda^3}\;+\;
\nu\lambda^3\;=\;0\ee where L is the conserved angular momentum of
the configuration and therefore the conserved energy is

\be\label{eq10}
E\;=\;\frac{\dot{\lambda}^2}{2}\;+\;\frac{L^2}{2\lambda^2}\;+\;
\frac{\nu}{4} \lambda^4 \ee

When $L=0$ the collapsing membrane solution is obtained \cite{CT}.
When $L\neq 0$ the membrane can be stabilized at the minimun of
the effective potential.

%A simple illustration of the above factorization conditions is given by
%choosing the hermitian matrix A to be the identity $3\times3$
%complex matrix which implies $\mu=1$ and the three functions

%The N-H solution for spherical topology of the membrane with
%target space the five dimensional sphere $S^{5}$ in six
%dimensional real target space( or $C^3$) is found, if we choose
%the three functions $f_{i}$, $f_{i}=e_{i},(\nu=2)$ to be the
%mapping to the unit sphere given by

%\be\label{eq5}e_{1}=\sin\theta \; \cos\phi,\;\; e_{2}=\sin\theta
%\; \sin\phi,\;\; e_{3}=\cos\theta\ee
We now turn to the case of the toroidal membrane. The four
functions $f_{i}$ are chosen by Nicolai-Hoppe \cite{NH} to be

\be\label{torus}
\vec{f}\;=\;(\cos\sigma_{1},\;\sin\sigma_{1},\;\cos\sigma_{2},\;
\sin\sigma_{2})\ee The target space in this case is the seven
dimensional sphere $S^{7}$ in eight dimensional real target space
(or $C^{4}$).

We now consider a different class of rotating configurations,
which slightly generalize the Nicolai-Hoppe\cite{NH} and
Hoppe\cite{TRRH} solutions, discussed previously by
Taylor\cite{TRRH}and more recently in \cite{HSS,AFP}.

In the framework of spherical matrix branes\cite{TRRH} by working
in  a rotating basis of the $SU(2)$ generators in the
N-dimensional representation instead of external magnetic fluxes
discussed by Myers\cite{KT}, these configurations stabilize the
attractive force of N $D0$ branes. Below we present the continuous
membrane analog of the matrix rotating ellipsoid \cite{HSS} in the
complex target space $C^{3}$\cite{AFP}.

\begin{equation}
\label{ellipsoid} Z = \left( \begin{array}{c} Z_1
\\ Z_2
\\ Z_3
\end{array} \right)=
\left(
\begin{array}{ccc} e^{i\omega_1 t} & 0 & 0 \\0 & e^{i\omega_2 t} & 0 \\
0 & 0 & e^{i\omega_3 t}
\end{array}\right)\left( \begin{array}{c} R_1 e_1(\theta, \phi)
\\ R_2 \; e_2(\theta, \phi)
\\ R_3 \; e_3(\theta, \phi)
\end{array} \right)
\end{equation}
where the functions $e_{i}(\theta,\phi)$ are given by (\ref{eq5})
 and form an $SU(2)$ algebra.
\be
\left\{ e_i,e_j \right\}=-\epsilon_{ijk} e_k \ee

%The ansatz \ref{ellipsoid} is of the form

% \be\label{eq11} Z_i (t,\theta,\phi) \;=\; \zeta_i (t) e_i
%~~~~f_i=e_i\ee

 %and the corresponding equations of motion are

%\be\label{eq12} \ddot{\zeta_{1}}\;=\; - (|\zeta_{2}|^2
%\;+\;|\zeta_3|^2 ) \zeta_1\ee \beq\label{eq13}\ddot{\zeta_2} \;=\;
%- (\left|\zeta_1\right|^2\;+\;\left|\zeta_3\right|^2) \zeta_2\ee
%\beq\label{eq14} \ddot{\zeta_3} \;=\;-(\left|\zeta_1\right|^2
%\;+\;\left|\zeta_2\right|^2) \zeta_3\ee

%\be\label{eq15}= V^{-1}\;A\;V\;=\;A_D\;=diag
%(\omega_1,\omega_2,\omega_3)\ee

 %\be\label{eq16} Z_i =
%\lambda(t)e^{i\omega_{i} t} f_i ~~~~~~ i=1,2,3\ee
The solution represents a rotating elipsoid of fixed shape of
three different axes $\lambda_{i}=R_{i}\;\;\; i=1,2,3$ with their
respective angular frequencies given by:

 \be\label{eq freq}\omega_{1}^{2}=R_{2}^{2}+R_{3}^{2}\;,\;\omega_{2}^{2}=
 R_{1}^{2}+R_{3}^{2}\;,\;\omega_{3}^{2}=R_{1}^{2}+R_{2}^{2}\ee

The corresponding three $U(1)$ charges are given by

 \be\label{charges} Q_{1}=
 \omega_{1}R_{1}^{2}\;\;,\;\;Q_{2}=\omega_{2}R_{2}^{2}\;\;,\;\;
 Q_{3}=\omega_{3}R_{3}^{2}\ee
and the energy (2.16) can be cast in terms of the $Q_{i}$ only
given above. Its form determines the equilibrium parameters of our
membrane configuration.

\section{ New Toroidal Membrane Solutions and Stability}

We shall show now that our generalized set up in complex $C^{n}$
flat spaces can easily accomodate toroidal solutions. They
generalize anticipated toroidal solutions by Nicolai and
Hoppe\cite{NH,TRRH}. We may also add that there is a keen interest
in the mathematical literature regarding the various embeddings of
Tori in $C^{3}$\cite{hu}

A natural basis of functions on $T^2$ is

 \be\label{eq25}f_{\vec{n}}\;\;=\;\; e^{ i \vec{n}\cdot\vec{\sigma}}\ee

In this basis we have that

\be\label{feqm} \left\{f_{\vec{n}},f_{\vec{m}}\right\}\;=\;-
(\vec{n}\times\vec{m}) f_{\vec{n}+\vec{m}}\ee  where
$\vec{n}\times\vec{m}= n_1m_2- n_2m_1$.

This is the area preserving infinite dimensional symmetry of the
torus $T^2$ namely Sdiff($T^2$)\cite{FI}. From eq.($2.1$) the
factorized form  $ Z_{i}(t)= \zeta_{i}(t) e ^{i
\vec{n}_{i}\cdot\vec{\sigma}}$ gives the following equations of
motion for $\zeta_i$:

\bea\label{eq32} \ddot{\zeta_{1}}&=& - \zeta_1 ( m^2 |\zeta_3|^2 +
k^2 |\zeta_2|^2 )\nn\\ \ddot{\zeta_2} &=& - \zeta_2 ( l^2
|\zeta_3|^2 + k^2 |\zeta_1|^2 )\nn \\ \ddot{\zeta_3}&=& - \zeta_3
( m^2|\zeta_1|^2 + l^2 |\zeta_2|^2 )\eea

The factorization condition is automatically satisfied for any
triplet $\vec{n}_{i} \in Z^2, \;\;\;i,j=1,2,3$

\bea\label{eq26} &&\frac{1}{2}\left\{
f^{\star}_{\vec{n}_{j}},\left\{
f_{\vec{n}_{j}},f_{\vec{n}_{i}}\right\} \right\}+
\frac{1}{2}\left\{
f_{\vec{n}_{j}},\left\{f^{\star}_{\vec{n}_{j}},f_{\vec{n}_{i}}\right\}\right\}=
-\nu_{ij} f_{\vec{n}_{i}}\\&& \nu_{ij}\;=\;(\vec{n}_{i} \times
\vec{n}_{j} )^2 \eea where

 \be\label{neqm} k^2\;=\;(\vec{n}_{1}\times\vec{n}_{2})^2 \;,\;
l^2\;=\;(\vec{n}_{2}\times\vec{n}_{3})^2\;,\;
  m^2\;=\;(\vec{n}_{3}\times\vec{n}_{1})^2 \ee

We get the toroidal solutions if we choose $ \zeta_i=R_i
e^{i\omega_{i}t},\;\;i=1,2,3$ with the angular frequencies being
related to the amplitudes as follows:

 \bea\label{eq29}
 \omega_{1}^{2}& = & k^{2} R_{2}^{2} + m^{2} R_{3}^{2}\\ \omega_{2}^2 &=&
 k^{2}  R_{1}^{2} + l^{2} R_{3}^{2}\\ \omega_{3}^{2}& =& m^{2} R_{1}^{2}+
 l^{2}R_{2}^{2}\eea

We observe that contrary to the case of the spherical membrane
solutions it is possible to choose $\vec{n}_{i}$ and for unequal
$R_{i},\;\;i=(1,2,3)$ such that all of the $\omega_{i}$ are equal.

In this case we observe that $|Z_i|^2=R_{i}^2\;\;\;i=1,2,3$ the
toroidal membrane moves in a $T^3= S^1\times S^1 \times S^1$ torus
with coordinates at any time t determined by the phases

\bea\label{eq31} \phi_{1}&=& \omega_{1} t + n_{1}^{1}\sigma_{1}+
n_{2}^{1}\sigma_{2}\nn\\ \phi_{2}&=&\omega_{2} t +
n_{1}^{2}\sigma_{1} + n_{2}^{2}\sigma_{2}\nn\\ \phi_{3} &=&
\omega_{3} t + n_{1}^{3} \sigma_{1} + n_{2}^{3}\sigma_{2} \eea

We have implicitly used  $U(1)^{3}$  as the remaining global
symmetry of eq.($3.3$) to fix the initial conditions for
$\phi_{i}$. With these phases, at any moment of time, the
embedding of $T^{2}$ inside $T^{3}$ can be expressed as
$Z_{i}\;=\;R_{i} e^{i\phi_{i}(t)}$. In order to visualize the
motion of  $T^{2}$  inside  $T^{3}$  it is helpful to write the
equation of the embedding in the periodic space of the phases
$\phi_{i},\;i=1,2,3$ which is a  $T^{3}$  of length  $2\pi$ as:

\be\label{phases} l\phi_{1}+m\phi_{2} + k\phi_{3}\;=\;(\omega_{1}
l + \omega_{2} m + \omega_{3} k) \;t \ee

This equation is derived by eliminating  $\sigma_{1},\sigma_{2}$
from eq($3.10$). We see that it is possible to have a time
periodicity if and only if  $\omega_i= (q_{i}/p_{i}) \omega$ where
the  $q_i$  and  $p_i$  are relative prime integers and $i=1,2,3$.
This is possible when the  $R_i$'s, which are the radii of  $T^3$
 take appropriate values. In this case the period is  $T= 2\pi
\frac{p}{\omega}$ and $p=lcm \left(p_1,p_2,p_3\right)$. We herein
denote ``$lcm$" the least common multiple. We also note that for
the special case of  $\vec{n}_{1}+\vec{n}_{2}+\vec{n}_{3}=0$  then
 $k^{2}=l^{2}=m^{2}$. In this case the dynamical equations of
motion  $(3.3)$  are similar to the spherical case and thus
eq.($3.11$) simplifies considerably.

The solution is invariant under the modular group $SL(2,Z)$ which
acts on $\vec{n}_{i}$. In other words the energy of the system
posesses an $SL(2,Z)$ degeneracy. We may also note that in this
case we have a ``twisted" $SU(2)$ Poisson algebra

\be\label{alge}
\left\{e^{î\vec{n}_{1}\vec{\sigma}},e^{i\vec{n}_{2}\vec{\sigma}}\right\}\;=
\; -(\vec{n}_{1}\times\vec{n}_{2})
e^{-i\vec{n}_{3}\vec{\sigma}}\ee and cyclic permutations.

In the following we shall demonstrate the radial stability of
these toroidal solutions. The linearized equations of motion
around $ Z_{i}^{c}\;=\; R_{i} \;e^{i\vec{n}_{i}\cdot\vec{\sigma}
+i \omega_{i}t }$ are easily found. The variations for radial
excitations  $ \delta\zeta_i\;=\; \zeta_{i}- \zeta_{i}^{c}$
satisfy the corresponding linear equation which can be obtained
directly from eq.$(3.3)$. For example for the case $(i=1)$:
% \be\label{eq27} \nu_{\vec{n}\vec{m}}=
%(\vec{n}\times\vec{m})^2\ee

%The eqs of motion thus finally become

%\be\label{eq28} \ddot\zeta_i\;=\; - \left ( \sum _{j=1}^3
%(\vec{n}^j \times \vec{n}^i)^2 |\zeta_j|^2\right) \zeta_i\ee

% For the specific case of $ \zeta_i(t)= R_{i}e^{i\omega_{i}t}$ we get

%\section{Stability of Breather Modes}

%We now proceed with the study of perturbations of the form
%$\zeta^{\prime}=\zeta + \delta\zeta$.They  satisfy the above eqs
%of motion. Let us take for example  the special case of
%$\zeta_{i}(t) = R_{i} e^{i\omega_{i}t}$.

%By plugging into eq.$(32)$ we get\bea\label{omega}
%\omega_{1}^{2}&=&k^{2}R_{2}^{2}+
%m^{2}R_{3}^{2}\nn\\ \omega_{2}^{2}&=& k^{2}R_{1}^{2}+l^{2}R_{3}^{2}\nn\\
%\omega_{3}^{2}&=& m^{2}R_{1}^{2}+l^{2}R_{2}^{2}\eea
% with the
%$\delta\zeta_i$s and their conjugates satisfying their associated
%eqs of motion. For example by focusing our attention on $\delta\zeta_{1}$ we observe that
% it satisfies

\bea\label{eq33} \delta\ddot{\zeta_{1}}&=&
-\delta\zeta_{1}\omega_{1}^{2} - \zeta^{c}_{1}[k^2
(\zeta^{c}_{2}\cdot\delta{\zeta_2}^{\star} +
\delta\zeta_{2}\cdot\zeta_{2}^{c\star})\nn\\& &+ m^{2}
(\zeta^{c}_{3}\cdot\delta\zeta_{3}^{\star}
 + \delta\zeta_{3}\cdot\zeta_{3}^{c\star})]\eea

In order to organize better our eigenvalue equation we go to the
body frame by introducing the transformation

\be\label{eq34} n_{i}= e^{-i\omega_{i}t}
\delta\zeta_{i}\;\;\;\;\;i=1,2,3\ee

By taking the appropriate time derivatives we eliminate the time
dependence of the coefficients of our eigenvalue equation through

\be\label{eq35} e^{-i\omega_{1} t} \delta\ddot{\zeta_1}=
\ddot{n}_{1}+ 2i\;\omega_{1}\;\dot{n}_{1}- \omega_{1}^{2}\;m \ee

We formulate the perturbation eqs. of motion for $\delta\zeta_1$
in terms of real and imaginary parts

\bea\label{eq36}\ddot{n}_{1R}-2\omega_{1}\dot{n}_{1I}&=&-2R_{1}\left(k^{2}
R_{2}n_{2R}+ m^{2}R_{3}n_{3R}\right)\\ \ddot{n}_{1I}+
2\omega_{1}\dot{n}_{1R}&=&0 \eea

As a consequence  we have that

\be\label{eq37} \frac{d\ddot{n}_{1R}}{dt}-2\omega_{1} \dot{n}_{1I}
= -2 R_{1}\left(k^{2} R_{2} \dot {n}_{2R} + m^{2} R_{3}
\dot{n}_{3R}\right)\ee Similar equations can be obtained for
$i=2,3$. We eliminate the imaginary part and in order to avoid
dealing with many indices we define

\be\label{eq37} \dot{n}_{iR}=u_{i}\;\;\;\;i=1,2,3 \ee

The eigenvalue eqs of motion take the form

\bea\label{eq38}\ddot{u_{1}} + (2\omega_{1})^2 u_{1}&=& -2R_{1}(
k^{2} R_{2}u_{2} + m^{2}R_{3}u_{3})\nn\\ \ddot{u_{2}}+
(2\omega_{2})^{2} u_{2}& = & -2R_{2}( k^{2}R_{1}u_{1} +
l^{2}R_{3}u_{3})\nn\\ \ddot{u}_{3}+ (2\omega_{3})^{2} u_{3} &=&
-2R_{3}(m^{2} R_{1}u_{1} + l^{2}R_{2}u_{2})\eea

We study the stability of the radial mode of the above equation.
We set $ u_{i}(t)= \xi_{i} \exp (i\lambda t) $ and obtain the
following matrix whose eigenvalues should be positive definite:

\be\label{eq39} M= \left[\ba{llll}  4\omega_{1}^{2} & 2
R_{1}R_{2}k^{2} & 2 R_{1}R_{3}m^{2}\\ 2R_{2}R_{1}k^{2}&
4\omega_{2}^{2}& 2R_{2}R_{3}l^{2}\\ 2R_{1}R_{3}m^{2}&
2R_{2}R_{3}l^{2}& 4\omega_{3}^{2}\ea\right] \ee

%We want to demonstrate positive definiteness and reality for the
%eigenvalues of M. We will utilize the fact that  if for a
%hermitian matrix A it is true for
 %every set of vectors $\xi_{i}$
%that  then its eigenvalues
%$\lambda_{i}$ are real and positive.

In order to demonstrate positive definiteness for the eigenvalues
of matrix M it is enough to show that $ \vec{\xi} \cdot M
\vec{\xi}
>0$ for every real vector $\vec{\xi}\in R^{3}$.

Indeed in our case we have that the corresponding  expression
takes the form

\bea \label{eq39} \vec{\xi}\cdot M\vec{\xi}&=& 4
k^{2}[(R_{2}\xi_{1} + R_{1}\xi_{2})^{2} - R_{1}R_{2}
\xi_{1}\xi_{2}]+ \nn \\& & 4m^{2}[ (R_{3}\xi_{1}+R_{1}\xi_{3})^{2}
-R_{1}R_{3}\xi_{1}\xi_{3}]\nn\\ & &+ 4l^{2}[( R_{3}\xi_{2}
+R_{2}\xi_{3})^{2} - R_{2}R_{3}\xi_{2}\xi_{3}]\eea which is
manifestly positive. With regard to the remaining three dimensions
($i=7,8,9$) the analysis for a general perturbation, not
necessarily radial, leads to bounded harmonic motion.

\section{Toroidal Bound States of N-D0 Branes}

We introduce $N\times N$ hermitian matrices $X_{i},\;\;i=1,...,6$.
In the complex notation we introduced previously we define

\bea\label{M1}Z_{1}&=& X_{1} +i X_{4}\nn\\ Z_2&=& X_2+i X_5
\\ Z_3&=& X_3 +i X_6\eea

The equations of motion take the following form

\be\label{M2} \ddot{Z_i}\;=\;
-\frac{1}{2}\left[Z_{j}^{\dagger},\left[Z_j ,
Z_i\right]\right]-\frac{1}{2}\left[Z_{j},\left[Z_{j}^{\dagger},
Z_{i}\right]\right] \ee

Gauss's Law equation of the constraint is given by

\be\label{M3} \left[\dot{Z}_{i},
Z_{i}^{\dagger}\right]\;+\;\left[\dot{Z}_{i}^{\dagger},
Z_{i}\right]\;=\;0 \ee

In analogy with (\ref{eq17}) of our membrane ansatz we make the
following matrix ansatz:

\be\label{M4} Z_{i}= \zeta_{i}(t) M_{i}\;\;\;i=1,2,3\ee where
$\zeta_i$ are the diagonals of a general nonsigular complex matrix
$\zeta_{ij}$ and $M_{i}$ is a general matrix that corresponds to
the $f_{i}$ functions of the membrane parameters. The constraint
equation (\ref{M3})
% $\left[M_{j},M_{j}^{\dagger}\right]=0\;\; \forall{j}$
is automatically satisfied with the NASC for the factorization of
time given by

\be\label{M5}
\frac{1}{2}\left[M_{j}^{\dagger},\left[M_{j},M_{i}\right]\right] +
\frac{1}{2} \left[M_{j}, \left[M_{j}^{\dagger},
M_{i}\right]\right]\;=\;\nu_{ji} M_{i}\ee for every $i$ and $j$,
see J.Hoppe in \cite{TRRH}. The $\zeta_i$ factors satisfy equation
(\ref{eq19}).

%\be \ddot{\zeta}_{i}\;=\;
%-\frac{1}{2}(\sum_{j}\left|\zeta_{j]\right|^{2}
%\nu_{ji}))\zeta_{i}\ee

As an example we consider the case of a spherical N-$D_{0}$ brane
bound state. In this case the N-$D_{0}$ branes constitute a fuzzy
sphere in six dimensions described by  $N\times N$ hermitian
matrices $M_{i}$. More specifically we consider the case
$M_{i}=J_{i}$ where $J_{i}$ are the three generators of the
$N=2j+1$ dimensional irreducible representation of $SU(2)$. By
plugging into both the equations of motion and that of the
constraint we find that $\nu_{ij}=2 ,\;\;i,j=1,2,3$. The analysis
is identical with the spherical membrane case
(\ref{eq20}-\ref{energy}) and (\ref{ellipsoid}-\ref{eq freq}).

We now proceed with the case of the torus $T^{2}$. We herein
consider $N\times N$ matrices

\be\label{M6} J_{\vec{n}}\;=\;\omega^{\frac{n_{1}n_{2}}{2}}
P^{n_{1}}Q^{n_{2}} \ee where $\omega \;=\; e^{\frac{2 \pi
i}{N}},\;\;\vec{n}=(n_{1},n_{2}) \in Z_{N}\times Z_{N}$

\be\label{M7} P\;=\;\left( \ba {ccccc} 0&0&\cdots&0&1\\
1&0&\cdots&0&0\\0&1&\cdots&0&0\\ \vdots& & & & \vdots
\\0&0&\cdots&1&0\ea\right)\;\;\; Q\;=\;\left( \ba{ccccc} 1& & &
&0\\ &\omega& & &
\\ & &\omega^{2}& & \\ & & &\ddots& \\0& & & &\omega^{N-1} \ea \right)\ee

From their definition $J_{\vec{n}}$, P and Q are unitary and
periodic matrices. They satisfy the conditions for a quantum
discrete torus \cite{FFZ};

\bea\label{M8}  Q\;P\;&=&\omega\;P\;Q \nn\\ P^{N}&=& I ,\;\;\;
Q^{N}=I\eea and

\be\label{M9} J_{\vec{n}}\;J_{\vec{m}}\;=\;
\omega^{-\left(\frac{\vec{n}\times\vec{m}}{2}\right)}\;\;
J_{\vec{n}+\vec{m}}\ee from which it also follows that

\be\label{M10} \left[ J_{\vec{m}},J_{\vec{n}}\right]\;=\;
-2\;\;i\; sin
\frac{\pi}{N}(\vec{m}\times\vec{n})\;\;J_{\vec{m}+\vec{n}}\ee

In the present case of the Matrix model we take as ansatz
$M_{i}=J_{\vec{n}_{i}},\;\;i=1,2,3$\cite{TRRH}. One can show that
in the limit $ N\rightarrow \infty,\;\;\omega\rightarrow 1$ we
recover the membrane parametrization  , namely $
J_{\vec{n}_{i}}\rightarrow f_{i}= e ^{i \vec{n}_{i}\vec{\sigma}}
\;\;i=1,2,3$\cite{FFZ}.

It is straightforward to check that the above ansatz gives us
%compute the double commutators for M in (\ref{M5}) and get the
%factorization of time condition as before for the case of the
%torus $T^2$ at hand

%\be\label{M11}
%\left[M_{j}^{\dagger},\left[M_{j},M_{i}\right]\right]+
%\left[M_{j},\left[M_{j}^{\dagger},M_{i}\right]\right]\;=\;4\left[sin
%\frac{\pi}{N}\left(\vec{n}_{j}\times\vec{n}_{i}\right)\right]^2\
$ \nu_{ji}\;=\; 4\left[sin\frac{\pi}{N}\left
(\vec{n}_{j}\times\vec{n}_{i}\right)\right]^2 $.

Our matrix ansatz (\ref{M4}) implies the following equations of
motion for  the $\zeta_{i}$:

\be\label{M12} \ddot{\zeta}_{1}\;=\; -4 \zeta_{1} \sum_{j\neq
i}\left[ sin \frac{\pi}{N}\left( \vec{n}_{j}\times
\vec{n}_{i}\right)\right]^{2} \left|\zeta_{j}\right|^{2}\ee

They translate into an identical set of equations for each
 $i=1,2,3$  as with the toroidal membrane case (\ref{eq32}) with
the proper identification for  $\nu_{ij}$ namely \bea\label{M13}
2\;sin \left(\frac{\pi}{N}k\right) \leftrightarrow k&=&
\vec{n}_{1}\times\vec{n}_{2}\nn
\\ 2\;sin\left(\frac{\pi}{N}m\right) \leftrightarrow m &=&
\vec{n}_{2}\times\vec{n}_{3} \\ 2\;sin\left(\frac{\pi}{N}l\right)
\leftrightarrow l&=& \vec{n}_{3}\times\vec{n}_{1}\nn\eea

By a complete similarity with the toroidal membrane if
$\vec{n}_{i}+\vec{n}_{2}+ \vec{n}_{3}\;=\;0 $ then $k=m=l$. The
algebra gets simplified and we get a twisted $SU(2)$ trigonometric
algebra $[J_{\vec{n}_{1}},J_{\vec{n}_{2}}]\;=\;-4\;i
\;sin\left(\frac{\pi}{N}k\right) J^{\dagger}_{\vec{n}_{3}}$ along
with their cyclic permutations. With this correspondence in mind
it can be observed that for the case of
$\zeta_{i}(t)=R_{i}\;e^{i\omega t}, \;\;i=1,2,3$ the angular
frequencies for the $N-D0$ Brane bound state ansatz admits a
similar dependence on their amplitudes. For the case $i=1$, for
example, we have that
 \be\label{M14} \omega_{1}^{2}=
 4\left[R^{2}_{2}\;\; sin^{2}\left(\frac{\pi}{N}k\right) +R_{3}^{2}
 \;\;sin^{2}\left(\frac{\pi}{N}m\right)\right]\ee

In this case the matrices $Z_{i}$ satisfy
 $Z_{i}Z_{i}^{\dagger}=R_{i}^{2},\;
 i=1,2,3$ namely the $N-D0$ bound
 state has the topology of a quantum $T^{2}$ torus embedded in a quantum
 $T^{3}$ torus with coordinates
 determined at any time $t$ by appropriate phases which are formally
 given by $(3.10)$.

 The stability analysis of our configuration subjected to radially
 symmetric fluctuations $ \delta\zeta_{i}= \zeta_{i}-\zeta_{i}^c$
 around the solutions $ \zeta_{i}^{c}= R_{i}\; e^{i\vec{n}_{i}\cdot\vec{\sigma}
 +i\omega t}$ proceeds identically with the membrane
 case and is given by eqs$(3.11-3.20)$ always taking properly into
 account the correspondence given by (\ref{M13}).
 Indeed we associate possitive numbers to
 positive numbers the positive definiteness of the eigenvalues
 of the matrix M of radial fluctuations does not get modified.

\section{Conclusions}

We have complexified the membrane and Matrix model dynamical
equations in even dimensional flat real spaces simplifying the
description of the existing known solutions as well as the search
for new solutions of rotating closed membranes and $N-D0$ brane
bound states correspondingly. Indeed we presented  constructions
of a new toroidal rotating membrane which is radially stable and
the motion of which is restricted in a $T^{3}$ torus. Its physical
interpretation is that of rotating black hole solutions of the
corresponding eleven dimensional supergravity \cite{TRRH}. This
interpretation extrapolates the gravitational duality of D-branes
as solutions of open string theory in flat spacetime with
boundaries to the case of supermembranes. This is easily
understood if we use as an intermediate step  the Matrix Model.
Indeed the latter can be used to connect supergravity with the
supermembrane theory. The stability consideration of classical
solutions is relevant to the quantization of supermembrane and
matrix model which has been recently in the focus of
attention.\cite{N}.

\section{Acknowledgements}
We thank  A.Kehagias  for useful discussions. This work was
supported by the EU under the TMR No CT2000-00122,00431. L.P.
acknowledges support from NCSR Demokritos where most of this work
was completed and by the C.Caratheodory Research Grant no. 2793
from  the U.of Patras. This work has also benefited from the
network ``Cosmology in the Lab" which is supported by the European
Science Foundation.

\bibliographystyle{prsty}

\bibliography{bibliog}

\end{document}